\input ./style/arxiv-general.cfg
\documentclass[aoas,MSNbibl,nameyear,dvips]{arximspdf}
\makeatletter
   \@ifpackageloaded{graphicx}{}{\usepackage{graphicx}}
\makeatother
\usepackage{algorithm,algorithmic}

\doi{10.1214/15-AOAS850}
\volume{9}
\issue{3}
\pubyear{2015}
\firstpage{1141}
\lastpage{1168}
\docsubty{FLA}

\makeatletter

\newcommand{\afrac}[2]{#1/(#2)}

\newcommand{\nfrac}[2]{#1/#2}
\renewcommand{\cite}{\citet}
\newcommand{\mN}{\mathcal{N}}
\newcommand{\iidsim}{\stackrel{\mathrm{i.i.d.}}{\sim}}
\makeatother

\begin{document}
\begin{frontmatter}

\title{Calibrating a~large computer experiment simulating radiative
shock hydrodynamics}
\runtitle{Calibrating a~large computer experiment}

\begin{aug}
\author[A]{\fnms{Robert B.}~\snm{Gramacy}\corref{}\thanksref{u1}\ead[label=e1]{rbgramacy@chicagobooth.edu}},
\author[B]{\fnms{Derek}~\snm{Bingham}\thanksref{u2}\ead[label=e2]{dbingham@stat.sfu.ca}},
\author[C]{\fnms{James Paul}~\snm{Holloway}\thanksref{u3}\ead[label=e3]{hagar@umich.edu}},
\author[C]{\fnms{Michael J.}~\snm{Grosskopf}\thanksref{u3}\ead[label=e4]{mikegros@umich.edu}},
\author[C]{\fnms{Carolyn C.}~\snm{Kuranz}\thanksref{u3}\ead[label=e5]{ckuranz@umich.edu}},
\author[C]{\fnms{Erica}~\snm{Rutter}\thanksref{u3}\ead[label=e6]{ruttere@umich.edu}},
\author[C]{\fnms{Matt}~\snm{Trantham}\thanksref{u3}\ead[label=e7]{mtrantha@umich.edu}}
\and
\author[C]{\fnms{R. Paul}~\snm{Drake}\thanksref{u3}\ead[label=e8]{rpdrake@umich.edu}}

\runauthor{R. B. Gramacy et al.}
\affiliation{University of Chicago\thanksmark{u1},
Simon Fraser University\thanksmark{u2} and University of
Michigan\thanksmark{u3}}
\address[A]{R. B. Gramacy\\
Booth School of Business\\
University of Chicago\\
Chicago, Illinois  60637\\
USA\\
\printead{e1}}
\address[B]{D. Bingham\\
Department of Statistics and\\
\quad  Actuarial Science\\
Simon Fraser University\\
Burnaby, British Columbia V5A 1S6\\
Canada\\
\printead{e2}}
\address[C]{J. P. Holloway\\
M. J. Grosskopf\\
C. C. Kuranz\\
E. R. Rutter\\
M. T. Trantham\\
R. P. Drake\\
Center for Radiative Shock\\\quad Hydrodynamics\\
University of Michigan\\
Ann Arbor, Michigan 48109\\
USA\\
\printead{e3}\\
\phantom{E-mail:\ }\printead*{e4}\\
\phantom{E-mail:\ }\printead*{e5}\\
\phantom{E-mail:\ }\printead*{e6}\\
\phantom{E-mail:\ }\printead*{e7}\\
\phantom{E-mail:\ }\printead*{e8}}
\end{aug}

%
\received{\smonth{10} \syear{2014}}
%
\revised{\smonth{6} \syear{2015}}

%
\begin{abstract}
We consider adapting a~canonical computer model calibration apparatus,
involving coupled Gaussian process (GP) emulators, to a~computer experiment
simulating radiative shock hydrodynamics that is orders of magnitude larger
than what can typically be accommodated. The conventional approach
calls for
thousands of large matrix inverses to evaluate the likelihood in an MCMC
scheme. Our approach replaces that costly ideal with a~thrifty take on
essential ingredients, synergizing three modern ideas in emulation,
calibration and optimization: local approximate GP regression,
modularization, and mesh adaptive direct search. The new methodology is
motivated both by necessity---considering our particular application---and
by recent trends in the supercomputer simulation literature. A~synthetic
data application allows us to explore the merits of several variations
in a~controlled environment and, together with results on our motivating
real-data experiment, lead to noteworthy insights into the dynamics
of radiative
shocks as well as the limitations of the calibration enterprise generally.
\end{abstract}

%
\end{frontmatter}

\section{Introduction}
\label{sec:intro}

Rapid increases in computational power have made computer\break  models (or
simulators) commonplace as a~way to explore complex physical systems,
particularly as an alternative to expensive field work or physical
experimentation. Computer models typically idealize the phenomenon being
studied, inducing bias, while simultaneously having more parameters than
correspond to known/controlled quantities in the field. Those extra ``knobs''
must be adjusted to make the simulator match reality. Computer model
\emph{calibration} involves finding values of such inputs, so that
simulations agree
with data observed in physical experiments to the extent possible, and
accounting for any biases in predictions based on new simulations.

Here, we are interested in computer model calibration for experiments on
radiative shocks. These are challenging to simulate because both hydrodynamic
and radiation transport elements are required to describe the physics. The
University of Michigan's Center for Radiative Shock Hydrodynamics
(CRASH) is
tasked with modeling a~particular high-energy laser radiative shock system.
The CRASH team developed a~code outputting a~space--time field that describes
the evolution of a~shock given specified initial conditions (the
inputs), and
has collected outputs for almost 27,000 such cases. The code has
two inputs involved in addressing known deficiencies in the mathematical
model, but which don't directly correspond to physical conditions. Our
goal is
to find values for these inputs, by calibrating the simulator to a~limited
amount of field data available from an earlier study, while simultaneously
learning relationships governing the signal shared between simulated
and field
processes in order to make predictions under novel physical regimes.

\citet{kennedy:ohagan:2001} were the first to propose
a~statistical framework
for such situations: a~hierarchical model linking noisy field measurements
from the physical system to the potentially biased output of a~computer model
run with the ``true'' (but unknown) value of any \emph{calibration parameters}
not controlled in the field.
The backbone of the framework
is a~pair of coupled Gaussian process (GP) priors for (a) simulator output
and (b)
bias. The hierarchical nature of the model, paired with Bayesian posterior
inference, allows both data sources (simulated and field) to contribute to
\emph{joint} estimation of all unknowns.

The GP is a~popular prior for deterministic computer model output
[\cite{sack:welc:mitc:wynn:1989}]. In that context, GP predictors are
known as
\emph{surrogate models} or \emph{emulators}, and they have many desirable
accuracy and coverage properties. However, their computational burden severely
limits the size of training data sets---to as few as 1000 input--output
pairs in
many common setups---and that burden is compounded when emulators are nested
inside larger frameworks, as in computer model calibration.
Consequently, new
methodology is required when there are moderate to large numbers of computer
model trials, which is increasingly common in the simulation literature
[e.g., \cite{kaufman:etal:2012,paciorek:etal:2013}].

Calibrating the radiative shock experiment requires a~thriftier apparatus
along several dimensions: to accommodate large simulation data, but
also to
recognize and exploit a~massive discrepancy between the relative sizes of
computer and field data sets. First, we modularize the model fitting
[\cite{liu:bayarri:berger:2009}] and construct the emulator using only the
simulator outputs, that is, ignoring the information from field data at that
stage. Unlike \citeauthor{liu:bayarri:berger:2009}, who argued for
modularization on philosophical grounds, we do this for purely computational
reasons. Second, we insert a~local approximate GP
[\cite{gramacy:apley:2014}] in place of the traditional GP emulator.
We argue
that the \emph{locality} of the approximation is particularly handy in the
calibration context which only requires predictions at a~small number
of field
data sites. Finally, we illustrate how mesh adaptive direct search
[\cite{AuDe2006}]---acting as glue between the computer model, bias
and noisy
field data observations---can quickly provide good values of calibration
parameters and, as a~byproduct, enough useful distributional
information to
replace an expensive posterior sampling.

The remainder of the paper is outlined as follows. Section~\ref
{sec:setup} describes the radiative shock application and our goals
in more detail. Section~\ref{sec:koh} then reviews the canonical calibration
apparatus with a~focus on limitations and remedies, including
approximate GP
emulation. Section~\ref{sec:method} outlines the recipe designed to
meet the
goals of the project.
Illustrations on synthetic
data are
provided in Section~\ref{sec:illus}, demonstrating proof of concept,
exploring variations and
discussing limitations. We return to the motivating example in
Section~\ref{sec:shock} equipped with a~new arsenal. The paper
concludes with a~brief
discussion in Section~\ref{sec:discuss}.

\section{Calibrating simulated radiative shocks}
\label{sec:setup}

The CRASH team is interested in studying shocks where radiation
from shocked matter dominates the energy transport and results in a~complex
evolutionary structure. These so-called radiative shocks arise in practice
from astrophysical phenomena (e.g., super-novae) and other high-temperature
systems [e.g., see \cite{McClarren2011,Drake2011}]. Our particular
work, here,
involves a~large suite of simulation output and a~small set of twenty field
observations from radiative shock experiments. Our goal is to calibrate the
simulator and to predict features of radiative shocks in novel settings.

\begin{figure}
\begin{tabular}{@{}c@{\quad\quad\quad}c@{}}

\includegraphics{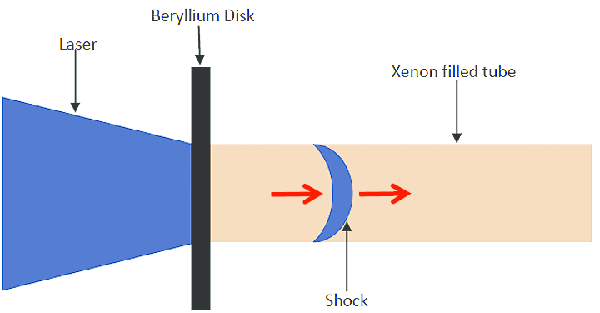}
&\includegraphics{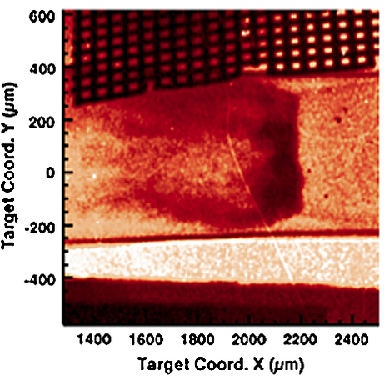}\\
\footnotesize{(a)}&\footnotesize{(b)}
\end{tabular}
\caption{\textup{(a)} Sketch of the apparatus used in the radiative shock
experiments. A~high-energy laser is used to ignite the
beryllium disk on the right, creating a~shock wave that travels through
the xenon filled
tube. \textup{(b)} Radiograph image of a~radiative shock
experiment.}\label{fig:expt}
\end{figure}

The field experiments were conducted at the Omega laser facility at the
University of Rochester [\cite{Boehly1997}]. A~high-energy laser was
used to
irradiate a~beryllium disk located at the front end of a~xenon (Xe) filled
tube [Figure~\ref{fig:expt}(a)], launching a~high-speed shock wave into the
tube. It is said to be a~radiative shock if the energy flux emitted by
the hot
shocked material is equal to or larger than the flux of kinetic energy into
the shock.
Each physical observation is a~radiograph image
[Figure~\ref{fig:expt}(b)], and the quantity of
interest for us is the \emph{shock location}: the
distance traveled at a~predetermined time.

The experimental (input) variables are listed in the first column of
Table~\ref{tab:inputs}, and the ranges or values used in the field
experiment (the
design) are in the final column. The first three variables specify the
thickness of the beryllium disk, the xenon fill pressure in the tube
and the
observation time for the radiograph image. The next four variables are
related to the geometry of the tube and the shape of the apparatus at its
front end. Most of the physical experiments were performed on circular shock
tubes with a~small diameter (in the area of 575 microns), and the remaining
experiments were conducted on circular tubes with a~diameter of 1150 microns
or with different nozzle configurations. The aspect ratio describes the shape
of the tube (circular or oval). In our field experiments the aspect ratios
are all 1, indicating a~circular tube. Our predictive exercise involves
extrapolating to oval shaped tubes with an aspect ratio of 2. Finally, the
laser energy is specified in Joules.

\begin{table}[b]
\caption{Design and calibration variables and input ranges for computer
experiment 1 (CE1) and 2 (CE2) and field experiments. A~single value
means that the variable was constant for all simulation runs}
\label{tab:inputs}
\begin{tabular*}{\textwidth}{@{\extracolsep{\fill}}lccc@{}}
\hline
\textbf{Input} & \textbf{CE1} & \textbf{CE2} & \textbf{Field
design} \\
\hline
\multicolumn{4}{@{}c@{}}{Design variables} \\
Be thickness (microns) & $[18,22]$ & 21 & 21 \\
Xe fill pressure (atm) &$[1.100,1.2032]$ & $[0.852,1.46]$ & $[1.032,1.311]$ \\
Time (nano-seconds) & $[5,27]$ & $[5.5,27]$ & 6-values in $[13, 28]$ \\
Tube diameter (microns) & 575 & $[575,1150]$ & $\{575, 1150\}$ \\
Taper length (microns) & 500 & $[460,540]$ & 500 \\
Nozzle length (microns) & 500 & $[400,600]$ & 500\\
Aspect ratio (microns) & 1 & $[1,2]$ & 1\\
Laser energy (J) & $[3600,3990]$ & & $[3750.0, 3889.6]$\\
Effective laser energy (J)\tabnoteref{TT1} & & $[2156.4,4060]$ & \\[6pt]
\multicolumn{4}{@{}c@{}}{Calibration parameters} \\
Electron flux limiter & $[0.04, 0.10]$ & 0.06& \\
Energy scale factor & $[0.40,1.10]$ &$ [0.60,1.00]$ & \\
\hline
\end{tabular*}
\tabnotetext{TT1}{The effective laser energy is the laser energy $\times$ energy
scale factor.}
\end{table}

Explaining the inputs listed in the remaining rows of Table~\ref{tab:inputs}
requires some details on the computer simulations. Two simulation
suites were
performed, separately, on super-computers at Lawrence Livermore and Los Alamos
National Laboratories, and we combine them for our calibration
exercise. The
second and third columns of the table reveal differing input ranges in
the two
computer experiments (denoted CE1 and CE2, resp.). Briefly, CE1
explores the input region for small, circular tubes, whereas CE2 investigates
a~similar input region, but also varies the tube diameter and nozzle geometry.
Both input plans were derived from Latin Hypercube samples
[LHSs, \cite{mckay:conover:beckman:1979}]. The thickness of the beryllium
disk could be held constant in CE2 thanks to improvements in
manufacturing in
the time in between simulation campaigns.

The computer simulator required two further inputs which could not be
controlled in the field, that is, two calibration parameters: the \emph
{electron
flux limiter} and the \emph{laser energy scale factor}. The electron
flux limiter
is an unknown constant involved in predicting the amount of heat transferred
between cells of a~space--time mesh used by the code. It was held
constant in
CE2 because in CE1 the outputs were found to be relatively insensitive
to this
input. 
The laser energy scale factor accounts for discrepancies between the amounts
of energy transferred to the shock in the simulations and experiments,
respectively. To explain, in the physical system the laser energy for
a~shock is recorded by a~technician. However, things are a~little more
complicated for the simulations. Before running CE1, it was felt that the
simulated shock would be driven too far down the tube for any specified laser
energy. Instead, the \emph{effective laser energy}---the laser energy actually
entered into the code---was constructed from two input variables, laser energy
and a~scale factor. For CE1 these two inputs were varied over the ranges
specified in the second column of Table~\ref{tab:inputs}. CE2 used
effective laser energy directly.

\begin{figure}

\includegraphics{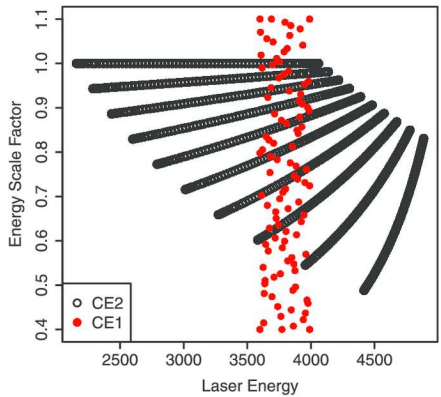}

\caption{Marginal design for laser energy and energy scale factor from
both experiments.}
\label{f:levesf}
\end{figure}

Our analysis uses both laser energy and the laser energy scale factor, which
is treated as a~calibration parameter. If the scale factor
``calibrates'' to
one, then there was no need to down-scale the laser energy in the first
experiment. Using both data sources requires reconciling the designs of the
two experiments. To that end, we expand the CE2 design by gridding
values of
laser energy scale factor and pairing them with values of laser energy deduced
from effective laser energy values from the original design. When
gridding, we
constrained the scale factors to be less than one but no smaller than value(s)
which, when multiplied by the effective laser energy (in reciprocal),
imply a~laser energy of 5000 Joules. Under those restrictions, an
otherwise uniform
grid with 100 settings of the scale factor yields a~total of 26,458
input--output
combinations, combining CE1 and expanded CE2 sets, to use in the calibration
exercise. Figure~\ref{f:levesf} shows the design over laser energy and energy
scale factor.

Our overarching goals here are three-fold: (a) design a~calibration
apparatus that can cope with data sizes like those described above, check
that we understand its behavior in controlled settings (synthetic
data), and
determine how best to deploy it for our real data (exploratory analysis);
(b) determine the settings of the two-dimensional calibration parameter,
note if down-scaling was necessary in CE1, and gain an understanding
of the extent to which the field data are informative about settings for
either parameter; (c) obtain (via a~particular setting of the calibration
parameter) a~high-quality predictor for field data measurements in novel
input conditions. In Section~\ref{sec:pred} we describe
a~(distribution of)
input setting(s) of interest to the CRASH team, for which field data have
been collected, which we use to benchmark our own predictions. Since this
experiment is for an oval-shaped disk, the predictions rely heavily on the
computer model output to make an extrapolation, as the field training data
observations involved only circular disks.

\section{Elements of computer model calibration}
\label{sec:koh}

As explained above, the radiative shock experiment involves runs of
a~deterministic computer model $M$ at a~large set of inputs $N_M =
26\mbox{,}458$, and a~much smaller number $N_F = 20$ of observations
from a~physical or field
experiment $F$. In what follows we refer to the inputs shared by $M$
and $F$
as \emph{design variables}, and denote them by $x$. The remaining (two
in our
case) calibration parameters required for $M$ are labeled as $u$, so
that $M$
takes inputs $(x,u)$. A~primary goal is to predict the result of new field
data experiments, via $M$, which means first finding a~good $u$. Below we
outline the elements involved in such an endeavor, with the focus on limitations
and remedies.

\subsection{Hierarchical models and modularization}

Kennedy and
O'Hagan\break
(\citeyear{kennedy:ohagan:2001}, hereafter KOH) proposed a~Bayesian
framework for coupling $M$ and $F$. Let $y^F(x)$ denote a~field observation
under conditions $x$, and $y^M(x,u)$ the (deterministic) output of a~computer
model run under conditions $x$ and calibration inputs $u$. KOH
represent the
\emph{real} process $R$ as the computer model output at the best setting
of the
calibration parameters, $u^*$, plus a~discrepancy term acknowledging that
there can be systematic disagreement between model and truth. In symbols,
$y^R(x) = y^M(x, u^*) + b(x)$.\footnote{We choose $b(x)$ for the discrepancy
term and casually refer to it as ``bias'' throughout even though the actual
bias, $y^M(x, u^*) - y^R(x)$, which is a~property of $M$ not $R$, would
actually work out to $-b(x)$.} The field observations connect reality with
data:
%
\begin{equation}
\label{eq:kohmodel}\quad y^F(x) = y^R(x) + \varepsilon=
y^M\bigl(x, u^*\bigr) + b(x) + \varepsilon  \qquad\mbox{where }
\varepsilon \iidsim\mN\bigl(0, \sigma_\varepsilon^2\bigr).
\end{equation}
The unknowns are $u^*$, $\sigma_\varepsilon^2$ and the bias
$b(\cdot)$. KOH propose a~Gaussian process (GP) prior for $b(\cdot)$,
which we
review in detail in the following subsection. Known information or
restrictions on $u$-values can be specified via a~prior $p(u)$, or
otherwise a~default/uniform prior can be used. Reference priors are
typical for
$\sigma_\varepsilon^2$.

If evaluating the computer model is fast, then inference is made rather
straightforward using residuals between computer model outputs and field
observations, $y^F(x) - y^M(x, u)$, which can be computed at will for any
$u$ [\cite{higdon2004combining}]. However, running the computer model is
usually time consuming, as is indeed the case in our example. In such
situations it is useful to use an \emph{emulator} or \emph{surrogate
model} in
place of $y^M(\cdot, \cdot)$. An emulator is a~fitted model $\hat
{y}^M(\cdot,
\cdot)$ trained on a~set of $N_M$ simulations of $M$ run over a~design
of $(x,
u)$-input values. KOH recommend a~GP prior for $y^M$. Rather than
performing inference for $y^M$ separately, using just the $N_M$ runs as is
typical of a~computer experiment in isolation
[e.g., \cite{morris:mitchell:ylvisaker:1993}], they recommend inference
joint with $b(\cdot)$, $u$ and $\sigma_\varepsilon^2$ using both field
observations and runs of the computer model. From a~Bayesian
perspective this
is the coherent thing to do: infer all unknowns jointly given all
data.

It is also practical when the $M$ is \emph{very} slow, giving small $N_M$,
and, moreover, even a~small number $N_F$ of field data observations can be
highly informative about the emulator $\hat{y}^M(\cdot,
\cdot)$ in that setting. But, more generally, this approach is fraught with
computational challenges. Coupled $b(\cdot)$ and $y^M(\cdot,
\cdot)$ lead to parameter identification and MCMC mixing issues, and
emulation demands substantial computational effort in larger $N_M$ contexts,
even when applied in isolation. These challenges are all compounded under
coupling.

\cite{liu:bayarri:berger:2009} propose going ``back to basics'' by
fitting the
emulator $\hat{y}^M(\cdot, \cdot)$ independently, using only the $N_M$
simulations. Inference for the rest of the KOH calibration apparatus is still
joint, for all parameters given $\hat{y}^M$ and field data $y^F$. Their
argument for this so-called \emph{modularization} is philosophical, and
is a~response to previous work outlining how fully Bayesian joint inference
in the
KOH framework unproductively confounds emulator uncertainty with bias
discrepancy
[\cite{sant:will:notz:2003}]. Our justification for entertaining modularized
calibration is different: decoupling has computational advantages.
Since our
$N_M \gg N_F$, a~small amount of field data cannot substantively
enhance the
quality of the emulator obtained under joint inference. In other words, we
don't lose much by modularizing. However, despite simplifying many
matters, a~marginalized approach would still require large $N_M$
emulation for our
application, and is therefore no panacea.

\subsection{Gaussian process emulation and sparse/local approximation}
\label{sec:laGP}

Gaussian process (GP) regression is canonical for emulating computer
experiments [\cite{sant:will:notz:2003}]. The reasons are many, but,
as we
shall see, computational tractability is not one of them. Technically,
the GP
is a~prior over functions between $x\in
\mathbb{R}^p$ and $Y \in\mathbb{R}$ such that any finite collection of
$Y$-values (at those $x$'s) is multivariate normal (MVN). Therefore, it is
defined by a~mean vector $\mu$ and covariance matrix $\Sigma$, and these
values may be specified in terms of hyperparameters and $x$-values.
Homoskedasticity and stationarity are common simplifying assumptions in
emulator applications. Often $\mu$ is constant/zero and $\Sigma= \tau
^2 K$
has constant scale $\tau^2$ and correlations $K$ defined only in terms of
displacements $x - x'$.

Performing GP regression requires applying the same logic,
conditionally on
data $D_N = (X_N, Y_N) = ([x_1^\top, \dots, x_N^\top]^\top, [y_1,
\dots,
y_N]^\top)$. Given values of any hyperparameters, the predictive
distribution for
$Y(x)$ at new $x$'s is directly available from MVN conditionals. Integrating
out $\tau^2$ under a~reference prior
[see, e.g., \cite{gramacy:polson:2011}] yields a~Student-$t$ with
%
\begin{eqnarray}
 \label{eq:predgp}\mbox{mean}\quad \mu(x|D_N, K_N) &=&
k^\top(x) K_N^{-1}Y_N,
\\
 \label{eq:preds2}\mbox{and scale} \quad \sigma^2(x|D_N,
K_N) &=& \frac{\psi[K(x, x) - k^\top(x) K_N^{-1} k(x)]}{N},
\end{eqnarray}
and $N$ degrees of freedom, where $k(x)$ is the $N$-vector whose
$i$th component is $K_\theta(x,x_i)$, defining the correlation
function given hyperparameters $\theta$; $K_N$ is an $N\times N$
matrix whose
entries are $K_\theta(x_i, x_j)$; and $\psi= Y_N^\top K_N^{-1} Y_N$.
Inference for $\theta$ can proceed by maximizing (e.g.,
Newton-schemes based on derivatives of) the likelihood,
%
\begin{equation}
\label{eq:gpk} p(Y|K_\theta) = \frac{\Gamma[N/2]}{(2\pi)^{N/2}|K_N|^{1/2}} \times \biggl(
\frac{\psi}{2} \biggr)^{-\nfrac{N}{2}},
\end{equation}
or via the posterior $\propto p(Y|K_\theta)p(\theta)$ in Bayesian schemes.

Observe that prediction and inference (even sampling from the GP prior)
requires decomposing an $N \times N$ matrix to obtain $K_N^{-1}$ and $|K_N|$.
Thus, for most choices $K_\theta(\cdot,
\cdot)$ and point-inference schemes, data sizes $N$ are limited
to the low thousands. Bayesian approaches are even further limited, as orders
of magnitude more likelihood evaluations (and matrix inversions) are
typically required, for example, for
MCMC. Assuming stationarity can also sometimes be too restrictive, and
unfortunately relaxation usually requires even more computation
[e.g., \cite{paciorek:schervish:2006,ba:joseph:2012,schmidt:ohagan:2003}].

A~key demand on the emulator in almost any computer modeling context, but
especially for calibration, is that inference and prediction (at
any/many $x$)
be fast relative to running new simulations (at $x$). Otherwise, why bother
emulating? As computers have become faster, computer experiments have become
bigger, limiting the viability of standard GP emulation. Sparsity is
a~recurring theme in recent searches for emulators with larger capability
[e.g., \cite{haaland:qian:2012,sang:huang:2012,kaufman:etal:2012,eidsvik2013estimation}],
allowing decompositions of large covariance matrices to be either avoided
entirely, be built up sequentially, or be carried out using fast sparse-matrix
libraries.

In this paper we use a~recent sparse GP methodology developed by
\cite{gramacy:apley:2014}. They provide a~localized approach to GP
inference/prediction that is ideal for calibration, where the full
inferential scheme (either KOH or modular) only requires $\hat
{y}^M(x,u)$ for
$(x,u)$-values coinciding with field-data $x$-values, and $u$-values
along the
search path for $u^*$, as we describe in Section~\ref{sec:method}.
The idea is to focus expressly on the prediction problem at an input
$x$. In
what follows we use $x$ generically, rather than $(x,u)$ as inputs to
$\hat{y}^M$. The local GP scheme acknowledges that data input
locations in
$X_N$ which are far from $x$ have vanishingly small impact on the predictive
equations (\ref{eq:predgp})--(\ref{eq:preds2}). This is used as the basis
of a~search for locations $X_n(x) \subset X_N$ which minimize Bayesian
mean squared
prediction error (MSPE). The search is performed in a~greedy fashion, giving
an approximate solution to the local design problem, and paired with efficient
updates to the local GP approximation as new data points are added into
the local
design. Building a~predictor in this way, ultimately using
equations~(\ref{eq:predgp})--(\ref{eq:preds2}) with a~data subset
$D_n(x)$, can be
performed in $O(n^3)$, a~substantial savings if $n \ll N$.
Pragmatically, one
can choose $n$ as large as computational constraints allow.

\cite{gramacy:apley:2014} show empirically that these MSPE-based local designs
lead to predictors which are more accurate than nearest
neighbor---using the
nearest $X_N$ values to $x$---which is known to be suboptimal
[\cite{vecchia:1988,stein:chi:welty:2004}]. They also extend the scheme
to provide local inference of the correlation structure, and thereby
fit a~globally nonstationary model. All calculations are independent
for each $x$, so local inference and
prediction on a~dense set of $x \in
\mathcal{X}$ can be trivially parallelized,
accommodating emulation for designs of size $N=10^6$ in under an hour
[\cite{gramacy:niemi:weiss:2014}]. An implementation is
provided in an $\mathsf{R}$ package called \texttt{laGP}
[\cite{laGP}]. However, independent calculations for each
$x$---while providing for nonstationarity and parallelization---yield
a~discontinuous global predictive surface, which can present challenges
in our calibration context.

\section{Proposed method}
\label{sec:method}

What we propose is thriftier than KOH in three ways, and thriftier than the
modularized version in two ways: It (a) modularizes the KOH hierarchical
model; (b) deploys local approximate GP modeling for the emulator
$\hat{y}^M(x, u)$; and (c) performs maximum  a~posteriori (point)
inference for
$u$ via the induced fits for the bias $\hat{b}(x)$ under a~GP prior.
Given a~value for the calibration parameter, $u$, the rest of the scheme
involves a~cascade of straightforward Newton-style maximizing
calculations. Below we
describe an objective function which, when optimized, performs the desired
calibration, giving an estimated value $\hat{u}$, for $u^*$. We then discuss
how to predict $Y^F(x)$ at new $x$-values given $\hat{u}$ and the data.

\subsection{Calibration as optimization}
\label{sec:opt}

Let\vspace*{-1pt} the field data be denoted as $D^F_{N_F} = (X^F_{N_F}, Y^F_{N_F})$, where
$X^F_{N_F}$ is the design matrix of $N_F$ field data inputs, paired
with an $N_F$
vector of $y^F$ observations $Y_{N_F}^F$. Similarly, let $D^M_{N_M} =
([X^M_{N_M},\break U_{N_M}], Y^M_{N_M})$ be the $N_M$ computer model input--output
combinations with column-combined $x$- and $u$-design(s) and $y^M$-outputs.
Then,\vspace*{-1pt} with an emulator $\hat{y}^M(\cdot, u)$ trained on $D^M_{N_M}$, let
$\hat{Y}^{M|u}_{N_F} = \hat{y}^M(X^F_{N_F}, u)$ denote\vspace*{1pt} a~vector of $N_F$
emulated output $y$-values at the $X_F$ locations obtained under a~setting,
$u$, of the calibration parameter. With local approximate GP modeling, each
$\hat{y}^{M|u}_j$-value therein, for $j=1,
\dots, N_F$, is obtained independently (and in parallel) via
local sub-design $X_{n_M}(x^F_j, u) \subset[X^M_{N_M},U_{N_M}]$ and locally
inferred hyperparameters $\hat{\theta}_j \equiv
\hat{\theta}(D_{n_M}(x^F_j, u))$. The size of the local sub-design,
$n_M$, is
a~fidelity parameter. Larger $n_M$ values provide more faithful
(compared to a~full GP)
emulation at greater computational expense. Finally, denote the $N_F$-vector
of fitted discrepancies as $\hat{Y}_{N_F}^{B|u} = Y^F_{N_F} -
\hat{Y}_{N_F}^{M|u}$.

Given these quantities, the quality of a~particular $u$ can be measured
by the
joint probability density of observing $Y^F_{N_F}$ at inputs
$X^F_{N_F}$. We\vspace*{-1pt}
obtain this from the best fitting GP regression\vspace*{-1pt} model trained on data
$D^{B}_{N_F}(u) = (X^F_{N_F},\hat{Y}_{N_F}^{B|u})$, emitting estimator
$\hat{b}$ for the bias given $u$.\footnote{Note that $D^{B}_{N_F}(u)$ tacitly
depends on hyperparameters $\hat{\theta}_j$ since it is defined
through local
GP emulation.} Values of $u$ which lead to a~higher probability of observing
$Y^F_{N_F}$ under the GP prior for $b(\cdot)$, modeling the discrepancy
between computer model emulations and field data, are preferred. We therefore
suggest finding $\hat{u}$ to maximize that probability, while simultaneously
maximizing over the parameterization of $b(\cdot)$, via hyperparameters
$\theta_b$, by solving the following optimization problem:
%
\begin{equation}\label{eq:obj}
\hat{u} = \arg\max_u \Bigl\{ p(u) \Bigl[ \max
_{\theta_b} p_b\bigl(\theta_b |
D^{B}_{N_F}(u)\bigr) \Bigr] \Bigr\}.
\end{equation}
Here $p(u)$ is a~prior for $u$ and $p_b(\theta_b | \dots)$ is
a~shorthand for
our bias ``fit'' $\hat{b}$: the marginalized posterior under a~GP
prior with
lengthscale hyperparmeters $\theta_b$ and noise parameter
$\sigma_\varepsilon^2$. It is computationally feasible to use a~full, rather
than approximate, GP for $b(\cdot)$ since $N_F$ is small. The ``inner''
$\max_{\theta_b}$ can be performed using Newton-like methods with closed-form
derivatives with respect to the lengthscale~$\theta_b$. The
``outer'' $\max_u$ is discussed shortly.

\begin{algorithm}
\begin{algorithmic}[1]
\REQUIRE Calibration parameter $u$, fidelity parameter $n_M$,
computer data $D^M_{N_M}$,\\ and field data $D^F_{N_F}$.
\FOR{$j=1, \dots, N_F$}
\STATE$I \leftarrow\mbox{\texttt{laGP}}(x^F_j, u | n_M, D^M_{N_M})$
\hfill\COMMENT{get indicies of local design}
\STATE$\hat{\theta}_j \leftarrow\mbox{\texttt{mleGP}}(D^M_{N_M}[I])$
\hfill\COMMENT{local MLE of correlation parameter(s)}
\STATE$\hat{y}^{M|u}_j \leftarrow\mbox{\texttt{muGP}}(x^F_j |
D^M_{N_M}[I],
\hat{\theta}_j)$
\hfill\{predictive mean emulation
following\\\hfill equation~(\ref{eq:preds2})\}
\ENDFOR
\STATE$\hat{Y}_{N_F}^{B|u} \leftarrow Y^F_{N_F} - \hat{Y}^{M|u}$
\hfill\COMMENT{vectorized bias calculation}
\STATE$D_{N_F}^{B}(u) \leftarrow(\hat{Y}_{N_F}^{B|u}, X^F_{N_F})$
\hfill\COMMENT{create data for estimating $\hat{b}(\cdot)|u$}
\STATE$\hat{\theta}_b \leftarrow\mbox{\texttt{mleGP}}(D_{N_F}^{B}(u))$
\hfill\COMMENT{full GP estimate of $\hat{b}(\cdot)|u$}
\RETURN$\mbox{\texttt{llikGP}}(\hat{\theta}_n, D_{N_F}^{B}(u))$
\hfill\COMMENT{the objective value of the {\texttt{mleGP}} call above}
\end{algorithmic}
\caption{Calculating the $p_b(\theta_b | D^{B}_{N_F}(u))$ term in
equation (\protect\ref{eq:obj})}
\label{alg:obj}
\end{algorithm}

Algorithm \ref{alg:obj} represents the ``inner'' max portion of
(\ref{eq:obj}) in pseudocode for a~more detailed second look. In our
implementation, steps 1--5
in the
code are automated by applying 
a~wrapper routine in the \texttt{laGP} package, called \texttt{aGP},
which loops
over each element $j$ of the predictive grid, performing local design,
inference for $\hat{\theta}_j$ and subsequent prediction stages, in parallel
via \texttt{OpenMP}. 
With $N_F$ and $n_M$ small relative to $N_M$, the execution of the
``for''-loop is extremely fast. In our examples to follow
(Sections~\ref{sec:illus}--\ref{sec:shock}), we use a~local
neighborhood size of
$n_M =
50$. Steps 8--9 are implemented by functions of the
same names in the \texttt{laGP} package.

The GP model for $b(\cdot)$, fit in step 8, estimates a~nugget
parameter (in addition to lengthscale $\hat{\theta}_b$) to capture
the noise
term $\sigma_\varepsilon^2$ in (\ref{eq:kohmodel}), whereas the local
approximate ones used for emulation, in step 3, do not. For
situations where bias is known to be very small/zero, it is sensible to
entertain a~degenerate GP prior for $b(\cdot)$ with an identity correlation
matrix. In that case, step 8 in Algorithm \ref{alg:obj} is skipped
and step 9 reduces to evaluating a~predictive density under an
i.i.d.~normal likelihood with $\mu=0$, that is, only averaging over
$\sigma_\varepsilon^2$. Note that Algorithm \ref{alg:obj} works with log
probabilities for numerical stability, while equation (\ref{eq:obj}) is
represented
in terms of unlogged quantities.

\subsection{Derivative-free optimization of the calibration objective}
\label{sec:dfo}

We turn now to the ``outer'' $\max_u$ in (\ref{eq:obj}), thinking of the
``inner'' $\max_{\theta_b}$ as an objective which can be evaluated following
Algorithm \ref{alg:obj}. The discrete nature of independent local design
searches for $\hat{y}^M(x_j^F, u)$ ensures that this objective is not
continuous in $u$. In fact, as we illustrate in our empirical work, it
can look
``noisy,'' although it is in fact deterministic. This means that optimization
with derivatives---even numerically approximated ones---is fraught with
challenges. 
We opt for a~derivative-free approach [see, e.g., \cite
{cohn:scheinberg:vincente:2009}].

Specifically, we use an implementation of the mesh adaptive direct search
(MADS) algorithm [\cite{AuDe2006}] called \texttt{NOMAD} [\cite{Le09b}],
via an
interface for $\mathsf{R}$ provided by the \texttt{crs} package
[\cite{crs}]. MADS
proceeds by successive pairs of \emph{search} and \emph{poll} steps, trying
inputs to the objective function on a~sequence of meshes which are
refined in
such a~way as to guarantee convergence to a~local optima under weak
regularity conditions; for more details see \cite{AuDe2006}. Direct, or
so-called pattern search, methods such as these have become popular for many
challenging optimization problems where derivative information is
either not
available or where approximations to derivatives may lead to
unstable numerical behavior. We are not the first to use MADS/\texttt
{NOMAD} in
the context of computer modeling. \cite{macdonald:ranjan:chipman} used
it to
search for the smallest nugget, leading to numerically stable matrix
decompositions for near-interpolating GP emulation. Our use is novel in the
calibration context.

As MADS is a~local solver, \texttt{NOMAD} requires initialization.
We recommend choosing a~starting $u$-value from evaluations on a~small
random space-filling design,
however, in our experiments (e.g., Section~\ref{sec:illus}),
starting at the center of the space performs almost as
well.

\subsection{Predictions for field data}
\label{sec:pred}

Posterior predictive samples of $Y^F(x)|\hat{u}$, representing the empirical
distribution of field-data observations at a~novel $x$ given
a~calibrated computer model using $\hat{u}$, can be obtained by running
backward through the KOH model (\ref{eq:kohmodel}) with estimated quantities
$\hat{b}(x)$ and $\hat{y}^M(x, \hat{u})$. That is,
obtaining a~predictive sample at $x$ involves executing the following steps
in sequence:
%
\begin{eqnarray}
\label{eq:lpred}\quad\quad Y_M &\sim&\hat{Y}_M\bigl(x|\hat{\theta}(x)\bigr)
\quad\mbox{via local GP under equations
(\ref{eq:predgp})--(\ref{eq:gpk})}\nonumber\\[-8pt]\\[-8pt]
&&\hphantom{\hat{Y}_M\bigl(x|\hat{\theta}(x)\bigr)
\quad}\mbox{with data
$D_{n_M}(x)$,} \nonumber
\\
\label{eq:bsim}Y_b &\sim&\hat{b}(x|\hat{\theta}_b) \quad \mbox{via full GP
under equations (\ref{eq:predgp})--(\ref{eq:preds2}) with data
 $D_{N_F}^{\hat{b}}(
\hat{u})$,}
\\
\label{eq:sum}Y_F &=& Y_M + Y_b  \quad \mbox{combining computer
model, bias and noise.}
\end{eqnarray}
On the left, above, we abuse notation somewhat and let estimated
emulator and
bias processes ``stand in'' for their corresponding predictive equations.
Pointers to those equations are provided on the right. In an unbiased version,
the zero-mean Student-$t$ draws in (\ref{eq:bsim}) are equivalent to
GP ones
with nugget-augmented diagonal correlation matrix $K =
\mathrm{diag}(1+\sigma_e^2)$ with both scale $\tau^2$ and noise
$\sigma
_e^2$ terms
integrated out. Equation (\ref{eq:lpred}) reminds that local GP
emulation depends
on both local design and locally estimated lengthscales.

Again consider Algorithm \ref{alg:obj} for a~second look. Field prediction
involves first running back through steps 2--4 to obtain a~local
design and correlation parameter [implementing equation (\ref{eq:lpred})],
parallelized for potentially many $x$; then performing steps 7--9
using saved $D^{\hat{b}}_{N_F}$ and $\hat{\theta}$ from the optimization
[equation (\ref{eq:bsim})]. However, rather than evaluate a~predictive
probability, instead save the moments of the predictive density (step
9) at the new $x$ locations. These can then be combined with the computer
model emulation(s) obtained in step 4, thus ``de-biasing'' the
computer model output to get a~distribution for $Y^F(x)|\hat{u}$, that is,
undoing step 6. Ideally, the full Student-$t$ predictive density
would be used here, in step 4, leading to a~sum of Student-$t$ random
variables [equation (\ref{eq:sum})] for $\hat{y}^M(x,\hat{u})$ and
$\hat{b}(x)$
comprising $y^F(x)|\hat{u}$. However, if $N_F, n_M \geq30$ summing normals
suffices, meaning no sampling is necessary.

As a~sum of random samples from a~convolution of two GP predictive
distributions, the resulting field predictions account for many uncertainties,
arising from both noise observed in the field and from model quantities
estimated from both data sources. Still, it is important to clarify
that some
uncertainties are overlooked in this approach. The biggest omission is
uncertainty in $\hat{u}$. Monte Carlo alternatives to optimizing $u$,
such as
posterior sampling or the bootstrap, are always an option. But these might
not be good value considering identification issues known to plague KOH-style
calibration [\cite{loeppky:bingham:welch:2006}]. Our empirical work
shows that
predictions under $\hat{u}$ retain many desirable accuracy and uncertainty
attributes, despite (or in spite of) such clearly evident concerns.
When~$u^*$
is a~primary goal, we later show how \texttt{NOMAD} evaluations can be salvaged
to approximate a~(log) posterior surface, and that these largely agree
with a~much more expensive bootstrap alternative. Finally, deploying
point-estimates
(e.g., MAP) for lengthscales and other hyperparameters, like $\hat
{\theta}_b$
and $\hat{\theta}(x)$, is a~common ``Empirical Bayes'' practice. With
local GP
emulation, overlooking such uncertainties is one of many deliberate
acts of
pragmatism, including that of local design search. Since local GPs
overestimate uncertainty relative to full-data counterparts
[see, e.g., \cite{gramacy:haaland:2014}], a~measure of conservatism is
organically built in.

\section{Illustrations}
\label{sec:illus}

In this section we entertain variations on a~synthetic data-generating
mechanism akin to one described most recently by
\cite{goh:etal:2013}, who adapted an example from
\cite{bastos:ohagan:2009}. It uses two-dimensional field data inputs $x$,
and two-dimensional calibration parameters $u$, both residing in the unit
cube. The computer model is specified as follows:
%
\begin{equation}\label{eq:yM}
y^M(x, u) = \bigl(1-e^{-\afrac{1}{2x_2}} \bigr) \frac{1000 u_1 x_1^3 + 1900 x_1^2+2092 x_1+60}{100 u_2 x_1^3 + 500
x_1^2 + 4x_1+20}.
\end{equation}
The field data is generated as
%
\begin{eqnarray}\label{eq:yF}
y^F(x)  &=&  y^M\bigl(x, u^*\bigr) + b(x) + \varepsilon,\nonumber \\[-8pt]
\\[-8pt]
\mbox{where } b(x) &=& \frac{10x_1^2+4x_2^2}{50 x_1 x_2+10} \mbox{ and } \varepsilon\iidsim\mN\bigl(0,
0.5^2\bigr),\nonumber
\end{eqnarray}
using $u^* = (0.2, 0.1)$. We keep this setup, however, we diverge from
previous uses in the size and generation of the input designs, and the number
of field data replicates.

Our simulation study is broken into two regimes, considering biased and
unbiased variations, and is designed (i) to explore the efficacy of the
proposed approach; (ii) to investigate performance in different scenarios
(with/without bias, unreplicated and replicated experiments, etc.); and (iii)
to motivate alternatives for our real data analysis in Section~\ref{sec:shock}.
Both simulation regimes involve 100 Monte Carlo (MC) repetitions and proceed
as follows.

Each repetition uses a~two-dimensional LHS of size 50 (on the unit
cube) for
the field data design, with three variations on the number of replicates,
$\{1,2,10\}$, for each unique design variable setting, $x$, leading to $N_F
\in\{50, 100, 500\}$ random realizations of $Y^F$. The computer model design
begins with a~four-dimensional LHS of size 10,000. It is then augmented with
simulation trials that are aligned with the field data design. We take 10
points per input in the field data, differing only in the $u$-values:
the 500
total $(x_1, x_2)$-values are paired with a~two-dimensional LHS (also
of size
500) of $(u_1, u_2)$-values. Combining with the second LHS, this gives
$N_M =
10\mbox{,}500$ random $(x_1, x_2, u_1, u_2)$ locations for the deterministic
simulation of~$Y^M$.

In each MC repetition, a~\texttt{NOMAD} search for $\hat{u}$ is
initialized with
the best value found on a~maxmin design of size 20, which is obtained by
searching stochastically over a~two-dimensional LHS of size 200. Vague
independent $\mathrm{Beta}(2,2)$ priors on each component of $u$ discourage
the solver from finding solutions that lie on the boundary of the search
space. Finally, a~two-dimensional LHS of size 1000 is used to generate an
out-of-sample validation set of $y^F$ values without noise, that is,
$\varepsilon_{x} = 0$. Root mean-squared errors (RMSEs) and estimates
$\hat{u}$ of $u^*$ are our main metrics of comparison.

In addition to varying the number of replicates, our comparators include
variations on the calibration apparatus and emulation of $y^M$. For example,
we compare our local approximate modular approach (Section~\ref{sec:method})
to versions using the true calibration value, $u^*$, a~random value in the
two-dimensional unit cube, $u^r$, and combinations of those where $y^M$ is
used directly---that is, assuming free computer simulation, and thus bypassing
the emulator $\hat{y}^M$. On the suggestion of a~referee, we also
include GP
predictors derived from the field data $Y_{N_F}^F$ only, bypassing the
computer model and calibration parameter(s) entirely. Together, these
alternatives allow us to explore how the error in our estimates
decompose at
each level of the approximate modularized calibration.

\subsection{Unbiased calibration}

\begin{figure}

\includegraphics{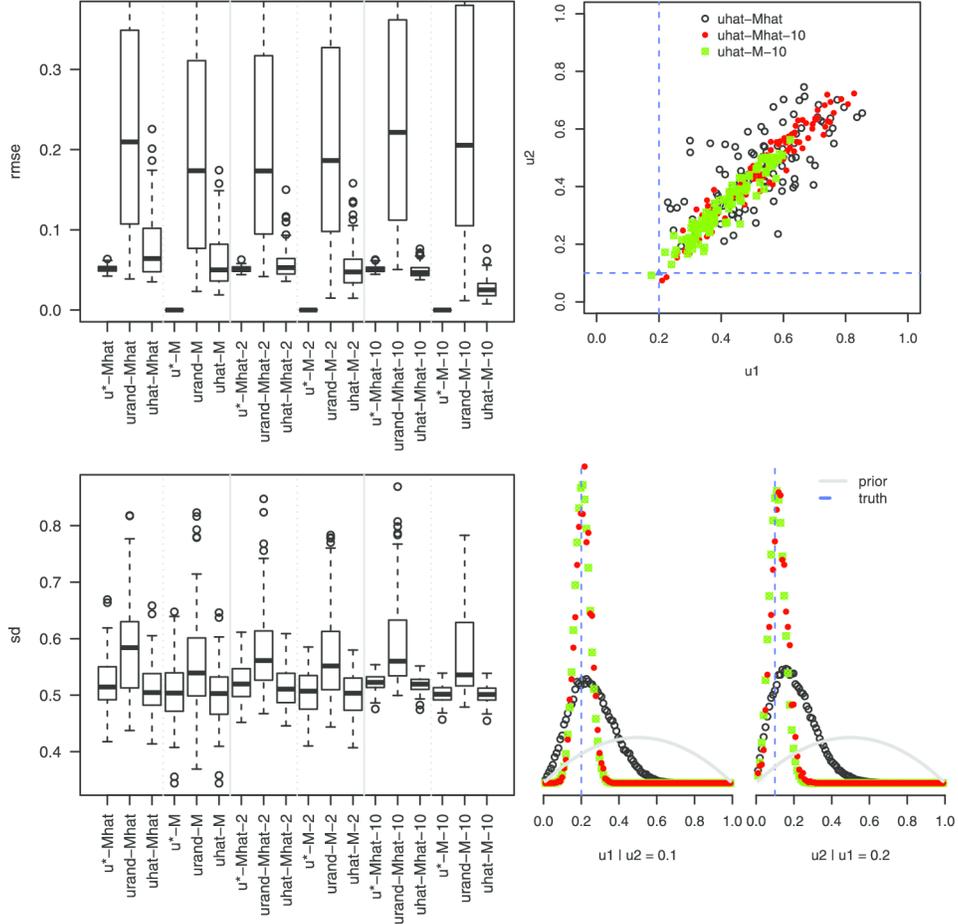}

\caption{Comparison on unbiased data, 100 MC replicates.
The  top left  panel shows RMSE to the true response on hold-out
sets, and
the bottom left  shows the corresponding standard deviations. The
 top right  panel shows three examples of the chosen calibration
parameter(s)
$\hat{u}$, and the  bottom right  shows 1-d density estimates
$\hat{u}_a$
conditional on the true value $u^*_b$ of the other coordinate.
True $u^*$ values are shown as dashed-blue lines, with
a~blue triangle positioned at their intersection. The boxplot axes and
scatter plot legend entries indicate if
$u$ is estimated (``uhat'') or if the true value is used
(``$u^*$'');
Field data
sets with 1, 2 and 10 replicates at each design location are shown,
arranged
in three groups of six along the $x$-axis in the  left  panels;
estimators
using $\hat{y}^M$ (``Mhat'') and $y^M$ (``M'') are grouped into three groups
of six. Whiskers of the ``urand'' boxplots are truncated to improve
visualization.}
\label{f:unbiased}
\end{figure}

Figure~\ref{f:unbiased} summarizes results from our first regime: generating
field data without bias, that is, setting $b(\cdot) = 0$ in
equation~(\ref{eq:yF}) and
fitting the model bias-free, that is, only estimating $\sigma
_\varepsilon^2$.
Consider the \emph{top left} panel first, which shows boxplots of RMSEs
arranged by numbers of replicates (three groups of six from left to right),
and then by the use of an emulator $\hat{y}^M$ or not (subgroups of three
within the six). Observe that a~random calibration parameter, $u^r$ (labeled
as ``urand,'' the middle boxplot in each group of three), gives poor
predictions of $y^F$. By contrast, using the correct $u^*$ with $y^M$ directly
(labeled ``\mbox{$\mathrm{u}^*$-M}'', fourth boxplot in each group of six), that is, not
emulating via
$\hat{y}^M$, leads to nearly perfect prediction. Contrasting with the
corresponding ``\mbox{$\mathrm{u}^*$-Mhat}'' boxplots (first in each group of six) reveals the
relative ``cost'' of emulating via $\hat{y}^M$ with~$u^*$. Together,
``urand''
and ``$\mathrm{u}^*$'' variations span the best and worst
alternatives. Distinctions
between the rest are more nuanced.

The third and sixth boxplots (from the left) show RMSEs obtained with
$\hat{u}$ via a~single field data replicate. RMSEs obtained under
$y^M$ or
$\hat{y}^M$ are very similar, with the former being slightly better.
This indicates that the local approximate GP emulator is doing a~good
job as a~surrogate for $y^M$. The story is similar for two replicates,
giving slightly
lower RMSEs (boxplots 9 and 12), as expected. Ten replicates (15 and
18) lead to greater differentiation between $y^M$ and $\hat{y}^M$ results,
implying more replicates provide a~more accurate and lower variance estimate
$\hat{u}$. Considering how bad things can get
(``urand''), all of the other estimates are quite good relative to the
best possible (``$\mathrm{u}^*$-M'' and ``$\mathrm{u}^*$-Mhat'').

The \emph{top left} panel does not include a~boxplot for the predictor
based on
fitting a~GP to the field data only---the comparator recommended by the
referee. We chose not to include these because of how they would adversely
affect the scale of the $y$-axis. The summary statistics (min, inter-quartile
range, and max) are as follows: $(0.44, 0.56, 0.73, 1.13)$ for one
repetition, $(0.31,
0.44, 0.59, 0.96)$ for two, and $(0.22, 0.30, 0.45, 0.97)$ for ten.
These are
pairwise dominated by every other comparator (with the same number of
replicates), including those based on random~$u^r$. Clearly, the computer
model/emulator is the key to good prediction.

The \emph{bottom left} panel shows estimated predictive standard deviations
(SDs) for each variation, whose corresponding RMSEs are directly above. SDs
are calculated by factoring in the predictive variances from both stages:
emulation uncertainty (if any), plus bias/noise components. The random
calibration parameter, $u^r$, gives the greatest uncertainty, which is
reassuring given its poor RMSEs. Uncertainties coming from $\hat{y}^M$ and
$y^M$ are very similar. 

The \emph{top right} panel shows estimated $\hat{u}$-values for three
representative cases. The others follow these trends and are omitted to
reduce clutter. In all three the $\hat{u}$-values found are along
a~straight line going through the true value $u^* = (0.2, 0.1)$. This
is the
case whether emulating with $\hat{y}^M$ or using $y^M$ directly,
although we
observe that when there are more replicates, or when $y^M$ is used directly,
the points cluster more tightly to the line and more densely near
$u^*$. We
conclude that there is a~ridge in the integrated likelihood for $u$, giving
equal density to combinations (e.g., in ratio) of $u_1$ and $u_2$ values.

This is confirmed in the \emph{bottom right} panel, which shows (MC average)
densities for one $u$-coordinate conditional on the true value of the other.
The pull of our prior, toward the center of the space, is visible in both
panes, but is far weaker when one of the coordinates is fixed. Further
simulation (not shown) reveals that, in this situation, weaker
$u$-priors move
estimates closer to the true $u^*$, however, uniform priors can yield
$\hat{u}$-values on the boundary, particularly near $u_2 = 0.2$. Also, observe
that the posterior evaluations appear ``noisy.'' This is an artifact of the
discrete nature of the local design search underlying $\hat{y}^M(x,
u)$. The
objective surface is in fact deterministic. Smoothly varying values of the
calibration parameter(s) may cause abrupt changes in the local design, and
lead to abrupt (if small) changes to local emulation and ultimately to the
maximizing posterior probabilities, motivating the \texttt{NOMAD} solver.

To wrap up with timing, we report that the most expensive
comparator (``uhat-Mhat-10'') took between 159 and 388 seconds,
averaging 232
seconds, over all 100 repetitions on a~16-core Intel Sandy Bridge 2.6
GHz Xeon
machine. That large range is due to variation in the number of
\texttt{NOMAD} optimization steps required, spanning 11 to 33,
averaging 18.

\subsection{Biased calibration}
\label{sec:illusbias}

\begin{figure}

\includegraphics{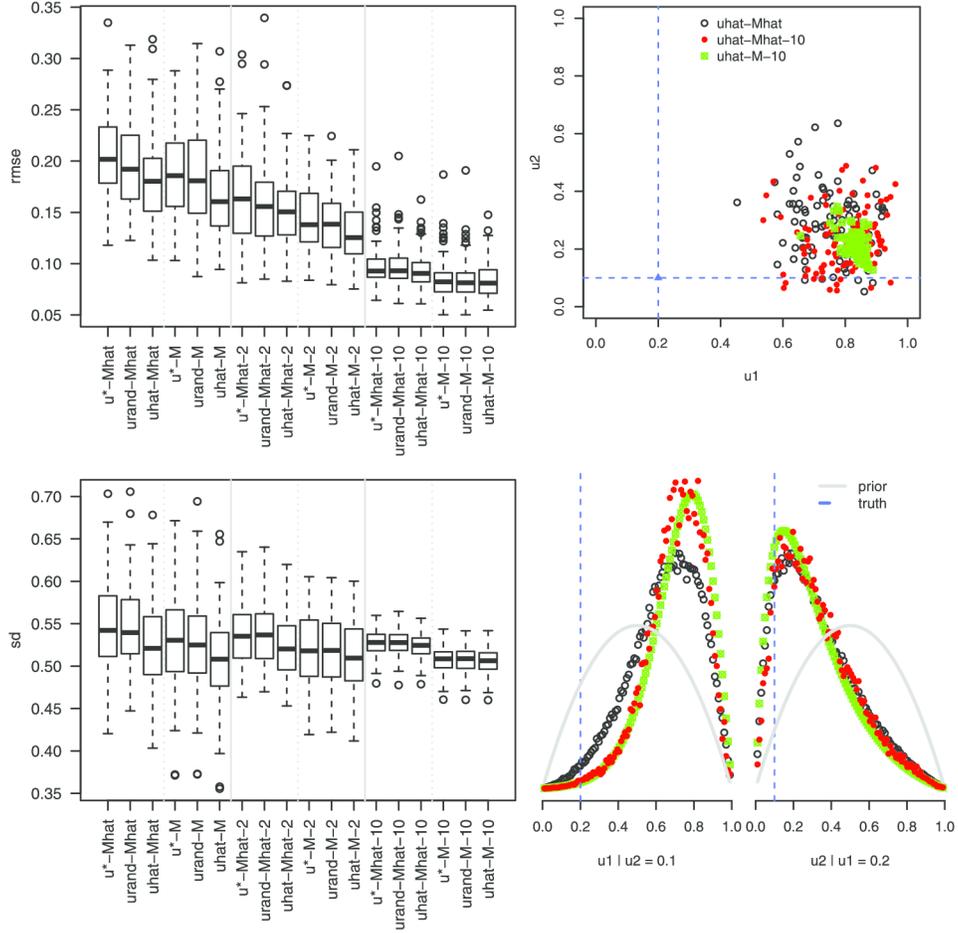}

\caption{Comparison on biased data, 100 MC replicates. The explanation
of the panels is the same as for Figure \protect\ref{f:unbiased}.}
\label{f:biased}
\end{figure}

Figure~\ref{f:biased} shows a~similar suite of results for the full, biased,
setup described in equations (\ref{eq:yM})--(\ref{eq:yF}), modeled with
a~GP prior on
$b(\cdot)$. At a~quick glance one notices the following: (1) the $\hat{u}$
estimates (\emph{top right}) are far from the true $u^*$ for all
calibration alternatives
considered; (2) the random setting $u^r$ isn't much worse than the other
options (\emph{top left}). Looking more closely, however, we can see
that the
$\hat{u}$ versions are performing the best in each section of the chart(s).
These are giving the lowest RMSEs (\emph{top left}) and the lowest SDs
(\emph{bottom left}). They are doing even better than with the true
$u^*$ setting.
So while we are not able to recover the true $u^*$, we nonetheless
predict the
field data better with the values we do find. Our modularized approximate
calibration method is excelling at one task, prediction of $y^F$,
possibly at
the expense of another, estimating $u^*$.

The explanation is nuanced. The bias (\ref{eq:yF}) is not well approximated
by a~stationary process, and neither is (\ref{eq:yM}) for that matter.
But our
fitted $\hat{b}$ assumes stationarity, so there is clearly a~mismatch with
(\ref{eq:yF}). The local approximate GP emulator does allow for
adaptivity of
correlation structure over the input space, and thus can accommodate a~degree
of nonstationary in the computer model (\ref{eq:yM}). That explains
why our
emulations were very good, but not perfect, in the unbiased case
(Figure~\ref{f:unbiased}). In this biased case, the full posterior
distribution,
inferring both full and local GPs, is using the flexibility of the joint
modeling apparatus to trade off responsibility, in effect exploiting
a~lack of
identifiability in the model, which is a~popular tactic in nonstationary
modeling (further discussion in Section~\ref{sec:discuss}). It is tuning
$\hat{u}$ to obtain an emulator that better copes with a~stationary
discrepancy, resulting in a~less parsimonious and larger magnitude
estimate of
$b$, but one for which $\hat{b}(\cdot) +
\hat{y}^M(\cdot, \hat{u})$ gives good predictions of $y^F(\cdot)$.
Meanwhile,
the local GP is faced with a~more demanding emulation task.

Again, we chose not to show boxplots for the field-data-only comparator
in the
figure because they would distort the $y$-axis scale. The
summary statistics are as follows: $(0.44, 0.57, 0.71, 1.13)$ for one
repetition, $(0.35,
0.46, 0.61,\break  1.23)$ for two, and $(0.20, 0.29, 0.47, 0.88)$ for ten.
These are
similar to the values obtained for the unbiased case, but it is
important to
note that they are not directly comparable since the data-generating
mechanisms are different---the former does not augment with
equation (\ref{eq:yF}).

Time-wise, the most expensive comparator (``uhat-Mhat-10'') took
between 538
and 1700 seconds, averaging 1049 seconds, over all 100 repetitions. The number
of \texttt{NOMAD} optimization steps was similar to the unbiased case, ranging
from 11 to 32, averaging 18. The main difference in computational cost
was compared to the unbiased case due to estimating the GP correlation structure
for $\hat{b}$, requiring $O(N_F^3)$ computations for $N_F = 500$.

\section{Calibrated prediction for radiative shocks}
\label{sec:shock}

We return now to our motivating example, having proposed a~thrifty framework
for calibration and explored its behavior in several variations on
a~representative benchmark problem. Our experimental setup for
calibration and
prediction is similar to the one described in Section~\ref{sec:illus}. In
particular, we again entertain both biased and unbiased alternatives, being
unsure about the extent of bias in the simulator relative to the field data.
One substantial distinction, however, between our synthetic data and
the radiative
shock experiment, concerns the input space and the local isotropy assumptions
underlying our local approximate GP emulator. This wasn't an issue in our
previous experiments since the inputs were in the unit cube, and the responses
(\ref{eq:yM})--(\ref{eq:yF}) varied by similar magnitude(s) within that range.

The radiative shock experiment involves a~larger (and disparate unit) input
space (Table~\ref{tab:inputs}), therefore, we augment biased and unbiased
variations with pairings of two different types of preprocessing of the
inputs. Our first type of preprocessing simply scales all inputs to lie in
the unit 10-cube, mimicking our synthetic experiment. We call this the
``isotropic'' case, since all input directions share a~common
lengthscale. In
the local GP emulator, $\hat{y}^M$, local isotropy does not preclude global
anisotropy or even nonstationarity. However, the discrepancy $\hat{b}$ has
global reach, so isotropy can be restrictive---however, with only
twenty field
data observations, isotropy has the virtue of parsimony.

In a~second version we rescale those inputs by a~crude estimate of the global
lengthscale obtained from small random subsets of the computer model
run data.
Specifically, we randomly sample 1000 elements of the full 26,458
design, in
100 replications, and save the maximum a~posteriori estimate of a~separable
lengthscale hyperparameter from a~Gaussian correlation function.
The distribution of those lengthscales is summarized in Table~\ref{t:scales}.
Observe that while some inputs (the middle ones: Xe pressure, aspect ratio,
nozzle length, taper length, tube diameter) might cope well with a~common
lengthscale, the analysis suggests others require faster decay. Be thickness,
time and electron flux limiter benefit from lengthscales roughly 4$\times$ shorter
than those above; laser energy and energy scale factor almost~2$\times$. We entertain
dividing the (already cube-scaled) inputs by square roots of median
lengthscales to circumvent the limits of isotropy in estimating both
$\hat{y}^M$ and~$\hat{b}$.

\begin{table}
\tabcolsep=0pt
\caption{Summary of estimated lengthscales from a~separable power correlation
function applied 100 times to a~random subsample of size 1000 from the full
26,458 design}
\label{t:scales}
\begin{tabular*}{\textwidth}{@{\extracolsep{\fill}}lcccccccccc@{}}
\hline
& & & & & & & & & \textbf{Elect} & \textbf{Energy} \\
& \textbf{Be} & \textbf{Laser} & \textbf{Xe} & \textbf{Aspect} & \textbf{Nozzle} & \textbf{Taper} & \textbf{Tube} & & \textbf{flux} & \textbf{scale} \\
& \textbf{thick} & \textbf{energy} & \textbf{press} & \textbf{ratio} & \textbf{length} & \textbf{length} & \textbf{diam} & \textbf{Time} &
\textbf{limit} & \textbf{factor} \\
\hline
25\% & 0.17 & 1.94 & 3.26 & 2.68 & 3.54 & 3.15 & 3.26 & 0.51 & 0.51 &
2.48 \\
50\% & 0.64 & 2.11 & 3.65 & 2.94 & 3.85 & 3.57 & 3.55 & 0.69 & 0.88 &
2.73 \\
75\% & 1.05 & 2.33 & 4.07 & 3.25 & 4.20 & 3.95 & 3.77 & 0.91 & 1.35 &
2.98 \\
\hline
\end{tabular*}
\end{table}

Finally, a~few other small changes from Section~\ref{sec:illus} are worth
noting. We initialize the search for $\hat{u}$, a~two-vector
comprising of
electron flux limiter and energy scale factor, with a~larger space-filling
design (of size 200 compared to 20). Since we are not performing
a~Monte Carlo
experiment with hundreds of repetitions, we can afford a~more conservative,
computationally costly, search. When estimating the discrepancy $\hat
{b}$, we
apply a~GP model to the subset of inputs which actually vary on more
than two
values in the field data (laser energy, Xe pressure, time). See the final
column in Table~\ref{tab:inputs}. We drop tube diameter, which has
only two
unique settings, however, the results aren't much changed when it is included.

\subsection{Exploratory analysis}
\label{sec:explore}

Before providing results based on a~full calibration, in the four variations
described above, we report on an exploratory analysis concentrated on
stressing aspects of the full framework---emulation, bias, calibration and
prediction---with the aim of gaining insight into what differences
might be
expected under those variations, if any.

\begin{figure}

\includegraphics{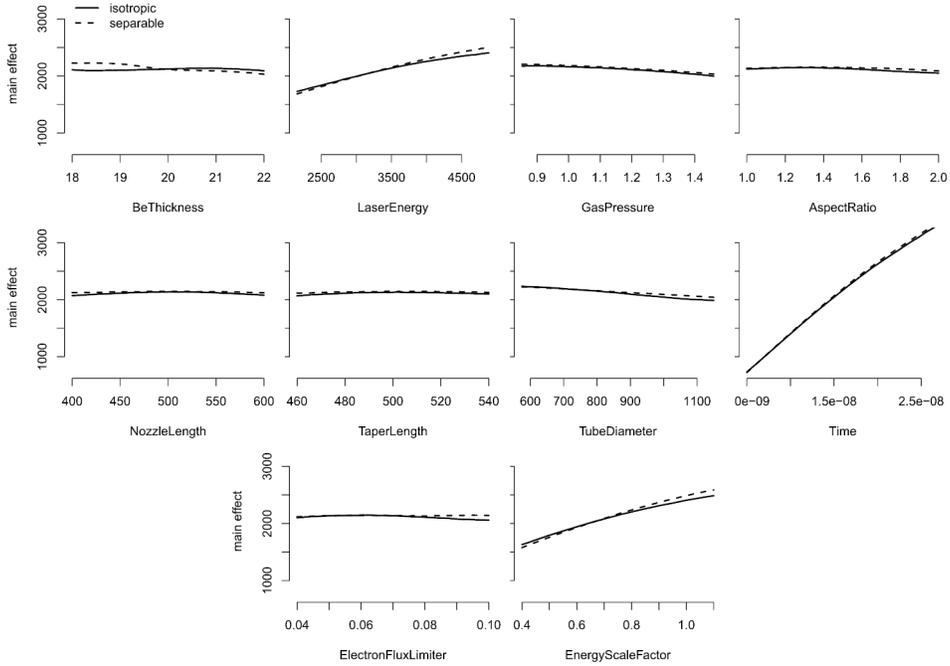}

\caption{Main effects plots for the emulated simulation runs.}
\label{f:SA}
\end{figure}

The first aspect is a~sensitivity analysis to see which inputs have
substantial impact on the response, with a~local GP emulator under both
isotropic and separable preprocessing regimes. Average main effect functions
are computed for each input [\cite{sobol1993}] and displayed in
Figure~\ref{f:SA}. Each panel of the plot gives the emulator response curve
for an
input, averaged over the remaining inputs. Observe in Figure~\ref
{f:SA} that
both preprocessing specifications give essentially the same results.
The most
influential inputs, marginally, are laser energy, time and laser energy
scale factor. The code is relatively less sensitive to the others, on average.
Foreshadowing somewhat, our prediction exercise in Section~\ref{sec:shockpred}
involves inputs with an aspect ratio of 2. Since there are no field
data runs
with that setting (see Table~\ref{tab:inputs}), even calibrated
predictions would be relying primarily
on the emulated computer model to make an extrapolation. The emulator
shows a~negligible effect for that input, so we can rest assured that
predictions in
this unsampled regime are not wildly different from where the models were
trained.

%
\begin{figure}

\includegraphics{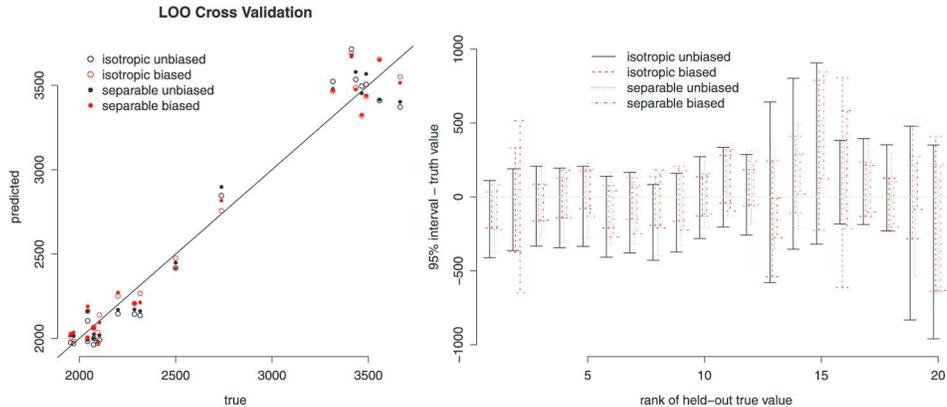}

\caption{Leave-one-out predictions for the radiative shock location
field data
versus true-values (left), and with error-bars after
subtracting out
the true value (right).}
\label{f:loo}
\end{figure}

We next report on a~leave-one-out study to asses the predictive ability
of the
four variations on our calibration methodology and gain confidence that
it is
capturing variability in the input space and between simulation and field
data. In turn, each of the twenty field observations is deleted, models
are fit
to the remaining observations and (all) simulations, and the deleted
observation is predicted.
The \emph{left} panel of Figure~\ref{f:loo} indicates that all four
methods are
performing well, with none obviously dominating the other in terms of
predictive means. Paired $t$-tests fail to detect differences in mean
predictive ability among all pairs of comparators. The \emph{right} panel
shows 95\% credible intervals from those predictions, after subtracting off
the true values. Here there may be some differences between the methods
visually. For example, the biased predictors seem to have the smallest
intervals, on average, which makes sense considering what we understand about
the data-generating mechanism. However, a~Bartlett test of unequal variances
fails to reject the null that all four predictors have the same variance.
This may be due to the small sample size of twenty.

\subsection{Model calibration}
\label{sec:shockcalib}

\begin{figure}

\includegraphics{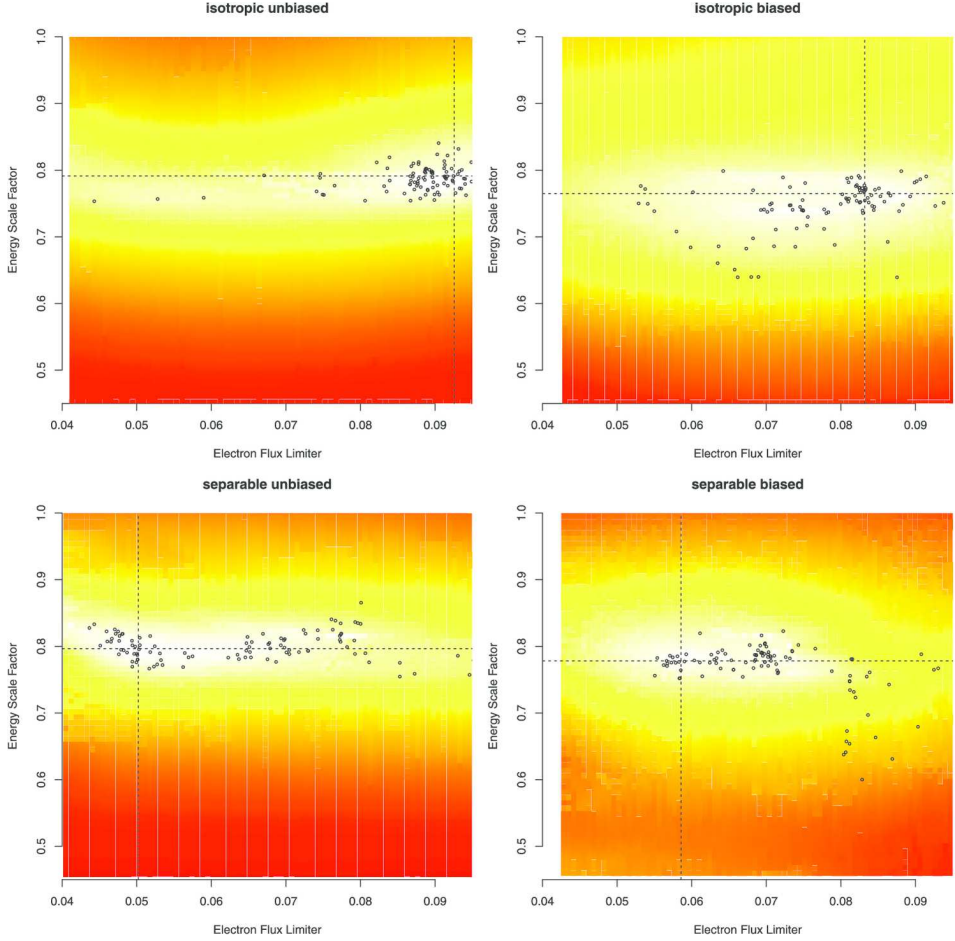}

\caption{Profile log-likelihood surfaces for the calibration parameters,
electron flux limiter and energy scale factor, in four setups.
Clockwise from
top left (MAP setting indicated by intersecting lines): isotropic unbiased;
isotropic biased; separable biased; separable unbiased. Open circles show
estimates obtained under parametric bootstrap resampling.}
\label{f:calibcrash}
\end{figure}

We turn now to a~full analysis of the calibration exercise in four variations.
The image plots in Figure~\ref{f:calibcrash} show the log posterior surface
interpolated from all evaluations of the objective (Algorithm \ref{alg:obj}),
combining the initial design and \texttt{NOMAD} searches. The
intersecting lines
indicate $\hat{u}$'s thus found, and the open circles are estimates obtained
under a~parametric bootstrap, discussed in more detail shortly. The unbiased
experiments took about 20 minutes to run on a~4-core hyperthreaded machine,
whereas the biased ones took fifteen. That ordering would seem paradoxical,
since the biased models have more quantities to estimate, however, the
\texttt{NOMAD} convergence was faster for the biased version,
requiring fewer
iterations navigate the posterior surface in search of $\hat{u}$.

Several observations are noteworthy. All four variations reveal that the
posterior surface is much flatter for the electron flux limiter than
for energy
scale factor, as expected. There is consensus on a~value of scale factor
between $0.75$ and $0.8$, meaning that scaling the laser energy in CE1 was
indeed helpful. The separable models, biased or unbiased, largely agree
on a~setting of the electron flux limiter, however, the isotropic versions
disagree with
that setting and disagree among themselves. We attribute this
divergence to
the scales estimated in preprocessing from Table~\ref{t:scales}.
Estimating a~bias adds fidelity to the model,
bringing estimates closer to those obtained in the separable version(s),
providing further illustration (augmenting the discussion in
Section~\ref{sec:illusbias}) of the dual role of the discrepancy
estimates in the
calibration framework.

\begin{figure}

\includegraphics{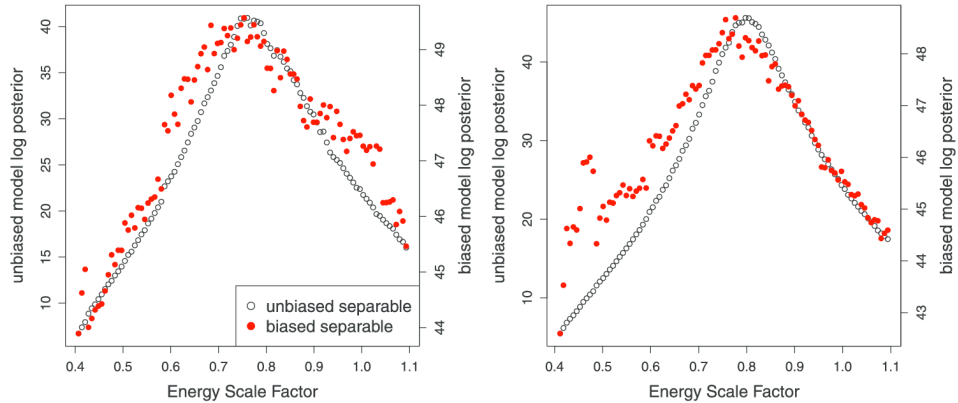}

\caption{Slice(s) of the profile log posterior surface over energy
scale factor with electron flux limiter fixed to its midway value in
the range:
isotropic on {left}; separable on {right}. In both plots, the
{left} axes show scale for the unbiased model, and the
{right} for the
biased one.}
\label{f:calibcrash:slice}
\end{figure}

As in our synthetic examples, observe that we obtain a~``noisy''
profile of the
log posterior in a~search for $\hat{u}$, although the objective is technically
deterministic. When the data are highly informative about good $\hat{u}$,
leading to a~peaked surface, the noise is negligible. However, when it is
flatter, the noise is evident. Figure~\ref{f:calibcrash:slice} shows
both cases via a~slice through the surface(s)
fixing the electron flux limiter at its midway value.
Being a~more flexible model, with weaker identification, the biased setup
yields a~much shallower log posterior surface. In the figure this is revealed
by the right-hand $y$ axes in both plots, compared to the left-hand ones.
Correspondingly, the red dots for biased posterior values are noisier.
The shallower and ``noisier'' surface may explain why \texttt{NOMAD} stopped
earlier---possibly prematurely---in the biased setup.


For a~second look at uncertainty in $\hat{u}$ we re-performed inference
on one
hundred parametric bootstrap re-samples of the field data observations
$Y_{N_F}^F$. See, for example, \citet{kleijnen:2014} for a~nice
review of
the bootstrap applied to models of simulation experiments. The resulting
estimates are shown as open circles in Figure~\ref{f:calibcrash}. Observe
that the bootstrap estimates agree with the heat plot depiction of
the posterior density, as interpolated from the \texttt{NOMAD}
samples. An
exception may be the separable unbiased case (\emph{bottom right}), which
contains a~dispersed cluster of lower energy scale factor estimates paired
with larger estimated electron flux limiter settings. It is important to
note that the bootstrap distribution would not, in general, be
identical to
the posterior surface. However, we draw comfort from their large
degree of
similarity in this example. The dual summaries of uncertainty in the
figure(s) suggest that the $\hat{u}$-values we estimated from the original
$Y_{N_F}^F$s are both representative (among open circles) and obtain high
probability (in light colored regions) under the posterior. If \texttt{NOMAD}
is indeed converging prematurely in the biased setup, due to the ``noise''
in the objective, the bootstrap results suggest it is still finding
highly probable $\hat{u}$ values.

\subsection{Prediction}
\label{sec:shockpred}

Next we make predictions on an interesting input setting provided to us
by the
CRASH team. The configuration is listed in the ``nominal settings''
column in Table~\ref{tab:design}.
 In past experiments, it was found that some of the desired input values were
not achieved for certain inputs when measured on the experimental apparatus
(i.e., in the field).
For example, the laser energy could be set
to 4000
joules, but a~laser energy of 3900 joules is what is observed. Our aim here
is to provide predictions for field data experiments before they are
run on
the apparatus. Therefore, for three of the variables the CRASH team
provided a~distribution over the inputs (third column in the table). In
the case
of Be
thickness, no variation was observed in past experiments, but as
a~conservative accounting of uncertainty, the input was sampled from a~uniform
distribution within manufacturing specifications. We were asked to propagate
these uncertainties through the calibrated predictive\vadjust{\goodbreak} model(s).

\begin{table}
\caption{Settings and distributions for the design variables in the 2012
experiments. The Be thickness is uniform over the specified range and the
Laser energy and Xe fill pressure are both normal with the specific
mean and
standard deviation}
\label{tab:design}
\begin{tabular*}{\textwidth}{@{\extracolsep{\fill}}lcc@{}}
\hline
\textbf{Input} & \textbf{Nominal value} & \textbf{Distribution} \\
\hline
\multicolumn{3}{@{}c@{}}{Design variables}\\
Be thickness (microns) & 21& $\operatorname{Unif}(20.5, 21.5)$ \\
Laser energy (J) & 3800& $\mathcal{N}(3800, 81.64)$ \\
Xe fill pressure (atm) &1.15 & $\mathcal{N}(1.15, 0.10)$ \\
Tube diameter (microns) & 1150 & \\
Taper length (microns) & 500 & \\
Nozzle length (microns) & 500 & \\
Aspect ratio (microns) & 2 & \\
Time (ns) & 26 & \\
\hline
\end{tabular*}
\end{table}

In this manner the exercise is one of \emph{propagation uncertainty
quantification} in the most basic sense: determining how uncertain inputs
filter to uncertain outputs. As discussed in Section~\ref{sec:pred},
our calibration is able to further account for some additional
estimation uncertainties, like from emulation, estimation of bias and
observation error $\sigma^2_\varepsilon$, but not others like $\hat{u}$
without further simulation (e.g., a~bootstrap). To clarify, the scheme used
here is as follows: (i) sample an input $x$ according to Table~\ref
{tab:design}; (ii) sample from the predictive distributions for $y$ at
that $x$ given $\hat{u}$, as in Section~\ref{sec:pred}; (iii) repeat.
We note
that augmenting with iteration over bootstrap estimates of $\hat{u}$ produces
a~slightly larger spread, but these results are not shown here. The
goal of
this experiment is to explore how a~calibrated model (i.e., using one good
choice of $u$) predicts in a~small out-of-sample exercise.

\begin{figure}

\includegraphics{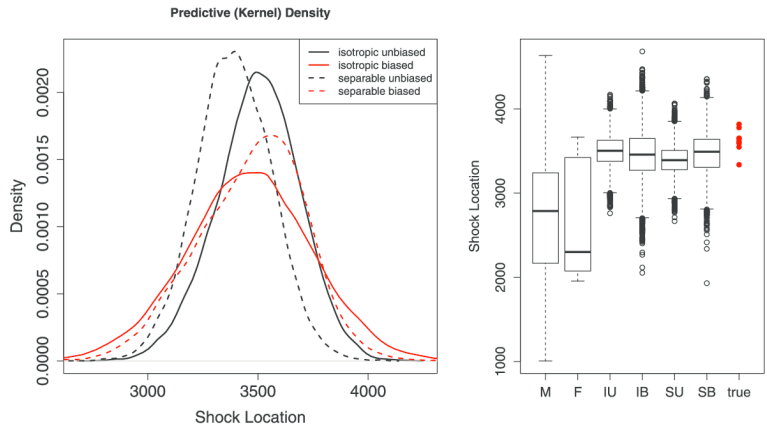}

\caption{Predictive densities for the 2012 experiments. The acronyms
IU, IB, SU, SB link boxplots on the {right} to the densities
shown plotted
on the {left}. M indicates the marginal computer model data; F indicates
the marginal distribution of the field data.}
\label{f:pred}
\end{figure}

Figure~\ref{f:pred}, focusing first on the \emph{left panel}, shows the
predictive distributions for our four variations. We first observe
that, on
the scale of the response marginalized over all inputs (roughly from
1000 to
4500), the predictive distributions are remarkably similar for all
methods, despite choosing different $\hat{u}$ for
the electron flux limiter. However, observe that estimating bias
leads to predictions (red densities) exhibiting a~greater degree of
uncertainty. Those models involve extra estimating steps and the random
values of the nominal settings from Table~\ref{tab:design} filter
through to mean and variance values for the estimated
bias. That the mode of the final distribution under the biased model
(dashed-red) is distinctly larger than the others, while at the same time
providing substantial spread for smaller values (but not larger ones---i.e.,
it is skewed toward the modes of the others), suggests that these predictions
are the most conservative. This squares well with an a~priori
preference for estimating bias and allowing separate lengthscales for each
input.

The \emph{right panel} shows a~boxplot version of the same distributions
alongside the output for a~field experiment subsequently performed at
the nominal
input settings in Table~\ref{tab:inputs}. From the plot we can see
that all
four distributions were quite accurate, showing greatest agreement with the
separable biased variation. We conclude that there is a~certain
robustness to
our calibration exercise(s), lending assurances to the methodology generally,
and to the predictions provided for the motivating application.

\section{Discussion}
\label{sec:discuss}

Motivated by an experiment from radiative shock hydrodynamics, we
presented a~new approach to model calibration that can accommodate
large computer
experiments, which are increasingly common in simulation-based applied work.
The cost of computation continues downward, with more and more
processor cores
being packed onto motherboards, and more nodes into computing clusters,
whereas the costs of field work remain constant (or possibly increasing).
Although the established, fully Bayesian KOH approach to calibration
has many
desirable features, we believe that it is too computation heavy to
thrive in
this environment. Something thriftier, retaining many of the salient features
of KOH, is increasingly essential.

Our method pairs local approximate Gaussian process (GP) emulation with
a~modularized approach to calibration, where the glue is a~flexible
derivative-free optimization method. The ingredients have been carefully
chosen to work well from an engineering standpoint. All software
deployed is
open source and available in $\mathsf{R}$. The extra subroutines we
developed have
been included in the \texttt{laGP} package on CRAN. During the time
that this
paper was under revision, we came across two works
[\cite{wong:storlie:lee:2014,damblin:etal:2015}] attacking computer
model calibration
leveraging similar themes: backing off of fully Bayesian aspects of
KOH, and
framing calibration as optimization. As we are, both of these papers are
motivated by pragmatism when it comes to devoting substantial computational
resources to quantities which are poorly identified. Our method is
unique in
its treatment of large-scale computer model emulation via local approximation,
and in providing open source software.

The biggest drawback of our approach is that it doesn't average over
uncertainty in the estimated calibration parameter $\hat{u}$. As demonstrated
in Figure~\ref{f:calibcrash}, output from the scheme can provide
insight into the
posterior for $u$, giving an indication of how robust a~particular
choice of
$\hat{u}$ might be. However, we do not provide a~method for sampling from
that distribution, as we believe that would require too much
computation to be
practical. As we demonstrate, a~parametric bootstrap is always an
option, which
is a~tack also taken by \cite{wong:storlie:lee:2014}. But in our real-data
example, it would seem that a~small amount of extra uncertainty comes at
the very high price of ${\sim}100\times$ greater computational effort.

We observed, as many have previously, that the calibration
apparatus can yield excellent predictions even when the estimated $\hat
{u}$ is
far from the true value. This can be attributed to the extreme flexibility
afforded by coupled nonparametric regression models, of which GPs are
just one
example, which further leverage an augmented design space: the calibration
parameters, $u$. Authors have recently exploited similar ideas toward
tractable nontstationary modeling. In the first case
\citet{ba:joseph:2012} proposed coupling GPs, and in the second
\citet{bornn:shaddick:zidek:2012} proposed auxiliary input variables.
We were surprised to discover that the KOH calibration model, preceding these
methods by nearly a~decade, effectively nests them: in the first without
auxiliary inputs, and in the second without bias.




%

\printaddresses
\end{document}